\documentclass[conference]{IEEEtran}
\IEEEoverridecommandlockouts
\usepackage{cite}
\usepackage{amsmath,amssymb,amsfonts}
\usepackage{algorithmic}
\usepackage{graphicx}
\usepackage{textcomp}
\usepackage{xcolor}
\usepackage[letterpaper, top=0.7in, right=0.575in, bottom=1.013in, left=0.673in]{geometry}

\usepackage{subfig}
\usepackage{hyperref}
\hypersetup{hidelinks,
	colorlinks=true,
	allcolors=black,
	pdfstartview=Fit,
	breaklinks=true}
\hypersetup{bookmarks=false}
\def\BibTeX{{\rm B\kern-.05em{\sc i\kern-.025em b}\kern-.08em
    T\kern-.1667em\lower.7ex\hbox{E}\kern-.125emX}}
\begin{document}

\title{PIB: Prioritized Information Bottleneck \\Framework for Collaborative Edge Video Analytics
}

\author{
    \IEEEauthorblockN{Zhengru Fang${}^{\star}$, Senkang Hu${}^{\star}$, Liyan Yang${}^{\star}$, 
Yiqin Deng${}^{\ddagger}$, Xianhao Chen${}^{\dagger}$, Yuguang Fang${}^{\star}$}
    \IEEEauthorblockA{$^\star$City University of Hong Kong, Hong Kong, China\\
    $^\ddagger$Shandong University, Jinan, China, 
    $^\dagger$The University of Hong Kong, Hong Kong, China\\
    Email: \{zhefang4-c, senkang.forest, liyanyang3-c\}@my.cityu.edu.hk,\\ yiqin.deng@email.sdu.edu.cn, 
    xchen@eee.hku.hk, my.fang@cityu.edu.hk}
    \vspace{-8mm}
}

\maketitle

\begin{abstract}

Collaborative edge sensing systems, particularly in collaborative perception systems in autonomous driving, can significantly enhance tracking accuracy and reduce blind spots with multi-view sensing capabilities. However, their limited channel capacity and the redundancy in sensory data pose significant challenges, affecting the performance of collaborative inference tasks. To tackle these issues, we introduce a Prioritized Information Bottleneck (PIB) framework for collaborative edge video analytics. We first propose a priority-based inference mechanism that jointly considers the signal-to-noise ratio (SNR) and the camera's coverage area of the region of interest (RoI). To enable efficient inference, PIB reduces video redundancy in both spatial and temporal domains and transmits only the essential information for the downstream inference tasks. This eliminates the need to reconstruct videos on the edge server while maintaining low latency. Specifically, it derives compact, task-relevant features by employing the deterministic information bottleneck (IB) method, which strikes a balance between feature informativeness and communication costs. Given the computational challenges caused by IB-based objectives with high-dimensional data, we resort to variational approximations for feasible optimization. Compared to TOCOM-TEM, JPEG, and HEVC, PIB achieves an improvement of up to 15.1\% in mean object detection accuracy (MODA) and reduces communication costs by 66.7\% when edge cameras experience poor channel conditions.
\end{abstract}


\section{Introduction}
Video analytics is rapidly transforming various sectors such as urban planning, retail analysis, and autonomous navigation by converting visual data streams into useful insights\cite{padmanabhan2023gemel}. When cameras are deployed for monitoring, they tend to produce vast amounts of video data constantly. There is often a requirement for quick analysis of these real-time streams\cite{fang2022age,fang2021stochastic}. Furthermore, numerous developing applications such as remote patient care\cite{wang2023contactless}, autonomous driving\cite{fang2024pacp}, and virtual reality heavily depend on efficient video analytics with minimal delay.

The increasing number of deployed smart devices require a computational paradigm shift towards edge processing. This approach involves handling data closer to its source, leading to several benefits compared to traditional cloud-based models. Specifically, this method significantly reduces latency, which is crucial in time-sensitive processing tasks like strokes and attacks. For example, as per the research conducted by Corneo \textit{et al}., utilizing remote cloud services for data processing may result in a 30\% increase in latency when compared to local data handling\cite{corneo2021much}. Moreover, the importance of privacy, particularly in regions with strict data protection laws such as the General Data Protection Regulation (GDPR), makes edge computing even more attractive\cite{marelli2018scrutinizing}. According to the Ponemon Institute, 60\% of companies express apprehension toward cloud security and decide to manage their own data onsites in order to mitigate potential risks\cite{Ponemon2021}.

However, the integration of edge devices into video analytics also brings in many significant challenges\cite{shao2023task}. The computational demands of deep neural network (DNN) models, such as GoogLeNet\cite{al2017deep}, which requires about 1.5 billion operations per image classification, place a substantial burden on the limited processing capacities of edge devices\cite{gao2024localization}. Additionally, the outputs from high-resolution cameras increase the communication load. For example, a 4K video stream requires up to 18 Gbps of bandwidth to transmit raw video data, potentially overwhelming the capacity of existing wireless networks\cite{yaqoob2020survey}.

The current communication strategies for integrating edge devices into video analytics ecosystems are not effective enough. One major issue is how to handle the computational complexity and transmission of redundant data generated from the overlapping fields of view (FOVs) from multiple cameras. In scenarios with dense camera deployments, up to 60\% of data can be redundant due to overlapping FOV, which unnecessarily overburdens the network\cite{jiang2020reinforcement}. In addition, these strategies often lack adaptability in transmitting tailored data features based on Region of Interest (RoI) and signal-to-noise ratio (SNR), resulting in poor video fusion or alignment. These limitations can negatively impact collaborative perception, sometimes making it less effective than single-camera setups.

In this paper, we aim to develop novel multi-camera video analytics by prioritizing wireless video transmissions. Our proposed Prioritized Information Bottleneck (PIB) framework attempts to effectively leverage SNR and RoI to selectively transmit data features, significantly reducing computational load and data transmissions. Our method can decrease data transmissions by up to 66.7\%, while simultaneously enhancing the mean object detection accuracy (MODA) compared to current state-of-the-art techniques. This approach not only compresses data but also intelligently selects data for processing to ensure that only relevant information is transmitted, thus mitigating noise-induced inaccuracies in collaborative sensing scenarios. This innovation sets a new benchmark for efficient and accurate video analytics at the edge.


\section{System Model and Problem Formulation}
\label{sec:system_model}
\begin{figure}[t]
  \centering
  \includegraphics[width=0.45\textwidth]{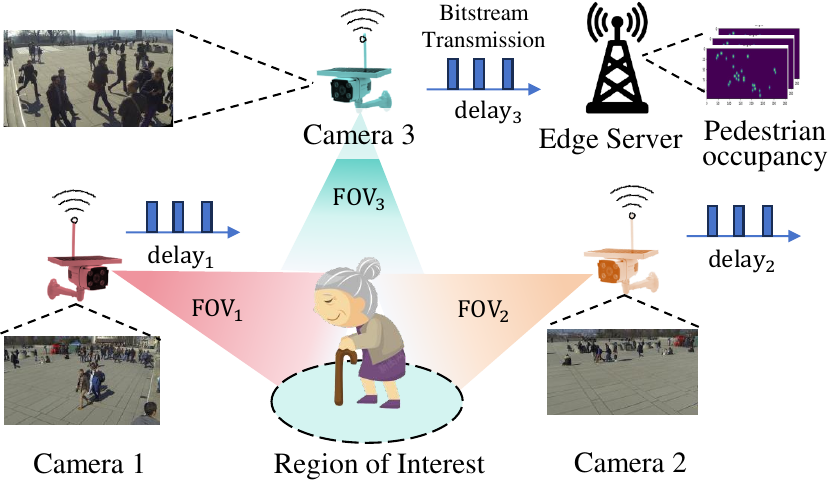}
  \caption{System model.}
  \label{fig:system model}
\end{figure}
As illustrated in Figure \ref{fig:system model}, our system comprises a set of edge cameras, denoted as \( \mathcal{K} = \{1, 2, \ldots, K\} \). These cameras are deployed across various scenes \( \mathcal{S} = \{s_1, s_2, \ldots, s_S\} \), each with a specific Field of View (FoV), \( \text{FoV}_k \), covering a subset of the total monitored area. The union of FoVs from all cameras covering a scene \( s \) ensures \( \bigcup_{k \in \mathcal{F}(s)} \text{FoV}_k \supseteq s \) for comprehensive surveillance. For example, in a high-density pedestrian environment, our goal is to facilitate collaborative perception for pedestrian occupancy prediction, under the constraints of limited channel capacity due to poor channel conditions.

\subsection{Communication Model}

Given the dense deployment and high device density, we adopt Frequency Division Multiple Access (FDMA) to manage the communication among cameras, defining the channel capacity \( C_k \) for each camera \( k \) using the SNR-based Shannon capacity:
\begin{equation}
C_k = B_k \log_2 \left(1 + \text{SNR}_k\right),
\end{equation}
where \( B_k \) is the bandwidth allocated to camera \( k \) and \( \text{SNR}_k \) is its signal-to-noise ratio. The transmission delay \( {d}_k \), critical for real-time applications, is calculated as:
\begin{equation}
{d}_k = \frac{D}{C_k},
\end{equation}
where \( D \) is the fixed data amount to be transmitted. This delay inversely correlates with \( C_k \), emphasizing the need for efficient bandwidth allocation and SNR optimization.

\subsection{Priority Weight Formulation}

Dynamic priority weighting is crucial in optimizing network resource allocation. We employ a dual-layer Multilayer Perceptron (MLP) to compute priority weights based on normalized delay and coverage:
\begin{equation}
p_k = \text{MLP}({d}_{\text{norm}, k}, \text{Coverage}_{\text{norm}, k}; {\Theta_{M}}),
\end{equation}
where \( p_k \) denotes the computed priority score for camera \(k\), and \({\Theta_{M}}\) represents the trainable parameters of MLP. This MLP's architecture, featuring two layers, allows for modeling the interactions between delay and coverage effectively. Besides, we have \( {d}_{\text{norm}, k} = \frac{{d}_k}{{d}_{max}} \) and \( \text{Coverage}_{\text{norm}, k} = \frac{{CO}_{k} - {CO}_{L}}{{CO}_{U} - {CO}_{L}} \). \( {CO}_{k} \) represents the coverage area provided by camera $k$ within the Region of Interest (RoI), with \( {CO}_{U} \) and \( {CO}_{L} \) denoting the upper and lower bounds of desired coverage, respectively.

To transform the raw priority scores into a usable format within the system, we apply a softmax function, which normalizes these scores into a set of weights summed to one:
\begin{equation}
w_k = \frac{e^{p_k}}{\sum_{j=1}^{K} e^{p_j}},
\end{equation}
where \( w_k \) signifies the priority weight for camera \(k\). This method ensures that cameras which are more critical, either due to higher coverage or due to lower delays, are given higher priority, thereby enhancing the decision-making capabilities and responsiveness of the edge analytics system.
\subsection{Video Feature Generation} 
In this paper, MVDet serves as the backbone network for multi-camera perception fusion\cite{hou2020multiview}. Firstly, edge cameras capture data and extract feature maps, which are subsequently encoded and sent via wireless channels to an edge server through base stations. Upon reception, the server decodes the data and uses a transformation matrix to perform coordinate transformations on the multi-view features, integrating them into a unified feature. The process concludes with the application of large kernel convolutions on the ground plane feature map, culminating in the final occupancy decision. This decision includes predicting the events of interest, e.g., pedestrian locations, enhancing the system's perception accuracy in multi-camera setups.

\subsection{Prioritized Information Bottleneck Formulation}
\label{sec: PIB}
In the context of information theory, the Information Bottleneck (IB) method seeks an optimal trade-off between the compression of an input variable $X$ and the preservation of relevant information for an output variable $Y$\cite{tishby2000information}. We formalize the input data from camera $ k $ as $ X^{(k)} $, similar to $ Z^{(k)} $ in extracted features, and the target prediction as $ Y^{(k)} $, corresponding to the population in the dataset $\mathcal{D} $. The aim is to encode $ X^{(k)} $ into a meaningful and concise representation $ Z^{(k)} $, resonating with the hidden representation $ z^{(k)} $ that captures the essence of multi-view content for prediction tasks. The classical IB problem can be formulated as a constrained optimization task:
\begin{equation}
\begin{aligned}
\max_{\Theta} \quad & \sum_{k=1}^{K}{I}\left(Z^{(k)};Y^{(k)}\right)\\
\text{s.t.} \quad  &{I}\left(X^{(k)};Z^{(k)}\right) \leq I_c,\quad (k=1,2,\cdots,K),
\end{aligned}
\end{equation}
where $I(Z^{(k)}, Y^{(k)})$ denotes the mutual information between two random variables $ Z^{(k)} $ and $ Y^{(k)} $. $\Theta$ represents the set of all learnable parameters in PIB framework, including ${\Theta_{M}}$ and the variational approximation in the following section. The mutual information is essentially a measure of the amount of information obtained about one random variable through the other random variable. $ I_c $ is the maximum permissible mutual information that $ Z^{(k)} $ can contain about $ X^{(k)} $. The objective is to ensure that $ Z^{(k)} $ captures the most relevant information about $X^{(k)}$ for predicting $ Y^{(k)} $ while remaining as concise as possible. Introducing a Lagrange multiplier $ \lambda $, the problem is equivalently expressed as:
\begin{equation}
\max_{\Theta} \quad R_{IB} =  \sum_{k=1}^{K}\left[{I}\left(Z^{(k)};Y^{(k)}\right) - \lambda \cdot {I}\left(X^{(k)};Z^{(k)}\right)\right],
\end{equation}
where $ R_{IB} $ represents the IB functional, balancing the compression of $ X^{(k)} $ against the necessity of accurately predicting $ Y^{(k)} $. Then, we extend the IB framework to a multi-camera setting by introducing priority weights to the mutual information terms, adapting the optimization for an edge analytics network:
\begin{equation}\label{eq:IB}
\min_{\Theta} \quad  \sum_{k=1}^{K} \left[ - I_{w}\left(Z^{(k)};Y^{(k)}\right) + \lambda I_{w}\left(X^{(k)};Z^{(k)}\right) \right],
\end{equation}
where the weighted mutual information terms are defined as: (1) $I_w\left(Z^{(k)};Y^{(k)}\right)=w_k\cdot I\left(Z^{(k)};Y^{(k)}\right)$ and $I_w\left(X^{(k)};Z^{(k)}\right)=e^{w^0-w_k}\cdot I\left(X^{(k)};Z^{(k)}\right)$. The non-negative value $w^0$ represents the maximum weight parameter $w_k$. 

The first term with linear weights \( I_w\left(Z^{(k)};Y^{(k)}\right) \) indicates the weighted mutual information between the compressed representation \( Z^{(k)} \) from camera \( k \) and the target \( Y^{(k)} \). Linear weighting by \( w_k \) ensures each camera’s influence is proportional to its priority, with higher \( w_k \) values increasing the emphasis on \( I\left(Z^{(k)};Y^{(k)}\right) \) in the objective function, emphasizing cameras that provide high-quality data for precise predictions. The second term with negative exponential weights \( I_w\left(X^{(k)};Z^{(k)}\right) \) measures the mutual information between the original \( X^{(k)} \) and its compressed \( Z^{(k)} \), scaled negatively by \( w_k \). This ensures exponential reduction in \( I\left(X^{(k)};Z^{(k)}\right) \) as \( w_k \) rises. Cameras with lower \( w_k \) undergo more aggressive data compression, optimizing bandwidth and storage without significantly impacting overall system performance. This weighting approach, chosen for this proof of concept, will be further explored with more general methods in future work.


\section{Methodology}\label{sec: method}

\begin{figure}[t]
  \centering
  \includegraphics[width=0.45\textwidth]{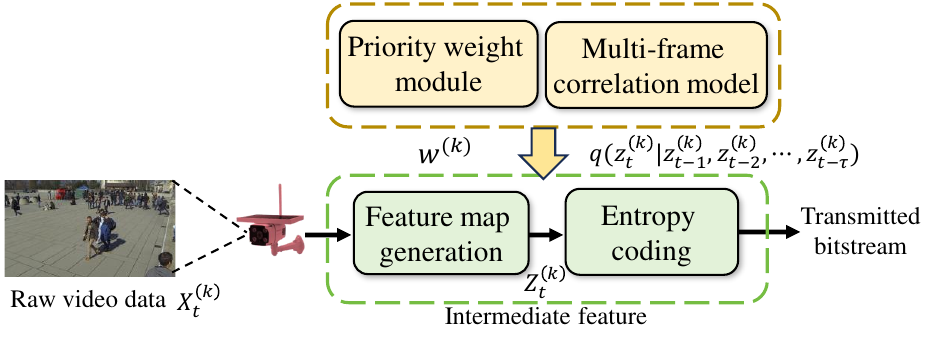}
  \caption{The procedure of video encoding.}
  \label{fig:encoder}
  \vspace{-3mm}
\end{figure}

\begin{figure}[t]
  \centering
  \includegraphics[width=0.45\textwidth]{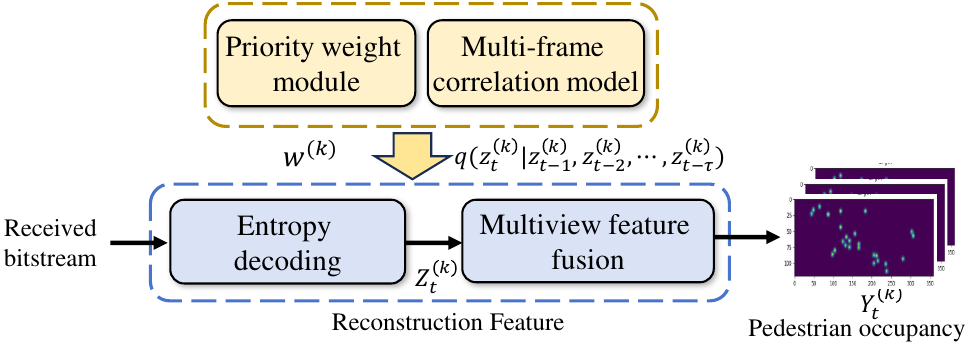}
  \caption{The procedure of video decoding.}
  \label{fig:decoder}
  \vspace{-3mm}
\end{figure}

\subsection{Architecture Summary}
In this subsection, we outline the workflow of our PIB framework, designed for collaborative edge video analytics. As depicted in Fig. \ref{fig:encoder}, the process starts with each edge camera (denoted by $k$) capturing raw video data $X_t^{(k)}$ and extracting feature maps. These cameras utilize priority weights $w_k$ to optimize the balance between communication costs and perception accuracy, adapting to varying channel conditions. The extracted features are then compressed using entropy coding and sent as a bitstream to the edge server for further processing. At the server (see Fig. \ref{fig:decoder}), the video features are reconstructed using shared parameters such as weights $w_k$ and the variational model parameters $q(Z_t^{(k)}|Z_{t-1}^{(k)}, ..., Z_{t-\tau}^{(k)})$. The server integrates these multi-view features to estimate pedestrian occupancy $Y_t$. This approach leverages historical frame correlations through a multi-frame correlation model to enhance prediction accuracy.

\subsection{Information Bottleneck Analysis}
The objective function of information bottleneck in Eq. (\ref{eq:IB}) can be divided into two parts. The first part is $-\sum_{k=1}^K w_k\cdot I\left(Z^{(k)};Y^{(k)}\right)$, which denotes the quality of video reconstruction by decoding at the edge server. The second part is $\lambda \sum_{k=1}^{K} e^{w^0-w_k}\cdot I\left(X^{(k)};Z^{(k)}\right)$, which denotes the compression efficiency for feature extraction. As it has been shown in the way that a decoder works, $p\left(Y^{\left(k\right)}|Z^{\left(k\right)}\right)$ can be any valid type of conditional distributions, but most often it is not smooth enough for straightforward calculation. Because of this complexity, it is highly challenging to directly work out the two mutual information components in Eq. (\ref{eq:IB}) and improve them. Accordingly, we adopt the variational approach\cite{alemi2016deep}. This approach suggests that the decoder is part of a simpler group of distributions called $Q$. We then search for a distribution $q\left(Y^{\left(k\right)}|Z^{\left(k\right)}\right)$ within this group that is most similar to the best possible decoder distribution, using the KL-divergence to measure the closeness. 

As a proof of concept, we first focus on deriving a lower bound for the mutual information \(I\left(Z^{\left(k\right)};Y^{\left(k\right)}\right)\) based on alternative probability distributions. We start with the standard definition of mutual information\footnote{For simplicity, we omit the exponent of ${\left(k\right)}$ when deriving the lower bound of $I\left( Z;Y \right)$.}:
\begin{equation}
\begin{aligned}
I\left( Z;Y \right) =\mathbb{E} _{p\left( Y,Z \right)}\left[ \log \frac{p\left( Y|Z \right)}{p\left( Y \right)} \right].
\end{aligned}
\end{equation}
We then introduce the Kullback-Leibler divergence (KL divergence), which is always non-negative and measures the efficiency of how the distribution \(q\left(Y|Z\right)\) approximates the true distribution \(p\left(Y|Z\right)\):
\begin{equation}\label{eq:KL}
D_{KL}\left[ p\left( Y|Z \right) ||q\left( Y|Z\right) \right] =\mathbb{E} _{p\left( Y|Z \right)}\left[ \log \frac{p\left( Y|Z \right)}{q\left( Y|Z \right)} \right] \ge 0,
\end{equation}
where the variational estimation method \( q\left(Y|Z\right) \) described utilizes a weighted exponential family variational distribution that is parameterized by neural network parameters \( \Phi \) and is designed to approximate the true conditional distribution \( p\left(Y|Z\right) \) while providing a computationally tractable lower bound for mutual information. Eq. (\ref{eq:KL}) leads to the inequality:
\begin{equation}
\mathbb{E} _{p\left( Y|Z \right)}\left[ \log p\left( Y|Z \right) \right] \ge \mathbb{E} _{p\left( Y|Z \right)}\left[ \log q\left( Y|Z \right) \right] ,
\end{equation}
where ${p\left( Y|Z \right)}$ can be replaced by ${p\left( Y,Z \right)}$. The relationship between the joint and conditional probabilities facilitates the simplification of the expression for mutual information:
\begin{equation}
\mathbb{E} _{p\left( Y,Z \right)}\left[ \log \frac{p\left( Y|Z \right)}{p\left( Y \right)} \right] =\mathbb{E} _{p\left( Z \right)}\left\{ \mathbb{E} _{p\left( Y|Z \right)}\left[ \log \frac{p\left( Y|Z \right)}{p\left( Y \right)} \right] \right\} .
\end{equation}

Building on the inequality established by the KL divergence (\ref{eq:KL}), we can express a lower bound for the mutual information:
\begin{equation}\label{eq: IZY}
\begin{aligned}
I(Z;Y) \geq \ \mathbb{E}_{p\left(Y,Z\right)}[\log q\left(Y|Z\right)] + H(Y),
\end{aligned}
\end{equation}
where \(H(Y)\) is the entropy of \(Y\), a constant that reflects the inherent uncertainty in \(Y\) independent of \(Z\). This formulation provides a computationally feasible lower bound for mutual information, crucial for applications in video analytics and other areas where direct computation of mutual information is hard or even infeasible.

As for the second part, we proceed by establishing an upper limit because of the complexity of directly minimizing the term $\lambda \sum_{k=1}^{K} e^{w^0-w_k}\cdot I\left(X^{(k)};Z^{(k)}\right)$. Recognizing that $H(Z^{(k)} | X^{(k)}) \ge  0$ from the properties of entropy, we can derive the following inequality:
\begin{small}
\begin{equation}\label{eq:hz}
\begin{aligned}
\lambda \sum_{k=1}^K{I_w}\left( X^{(k)};Z^{(k)} \right) \le \lambda \sum_{k=1}^K{\frac{H\left( Z^{(k)} \right)}{e^{w_k-w^0}}}\le \lambda \sum_{k=1}^K{\frac{H\left( Z^{(k)},V^{(k)} \right)}{e^{w_k-w_0}}},
\end{aligned}
\end{equation}
\end{small}
where we use the latent variables $V^{(k)}$ as the side information to encode the quantized feature and we have used $ H(Z^{(k)}, V^{(k)}) \geq   H(Z^{(k)})$. We begin by recognizing that the joint entropy $ H(Z^{(k)}, V^{(k)}) $ represents the communication cost. Then, we establish an upper bound by using the KL divergence non-negativity property:
\begin{equation}\label{ineq:second_part}
\begin{aligned}
H(Z^{(k)}, V^{(k)}) \leq & \mathbb{E}_{p(Z^{(k)}, V^{(k)})} \left[ -\log q(Z^{(k)}|V^{(k)}; \Theta_{con}^{(k)}) \right. \\
& \left. \times q\left(V^{(k)}; \Theta_{l}^{(k)}\right) \right].
\end{aligned}
\end{equation}
where \( \Theta_{con}^{(k)} \) and \( \Theta_{l}^{(k)} \) are the learnable parameters of the variational distributions \( q(Z^{(k)}|V^{(k)}; \Theta_{con}^{(k)}) \) and \( q(V^{(k)}; \Theta_{l}^{(k)}) \), respectively, which approximate the true distributions to minimize the communication cost while capturing the essential feature relations for inference. By taking Eq. (\ref{ineq:second_part}) into Eq. (\ref{eq:hz}), we obtain the upper bound for the second term in Eq. (\ref{eq:IB}), given by 
\begin{small}
\begin{equation}\label{ineq:second_part_final}
\begin{aligned}
I_w\left(X^{(k)};Z^{(k)}\right) \leq & \mathbb{E}_{p(Z^{(k)}, V^{(k)})} \left[ -\log q\left(Z^{(k)}|V^{(k)}; \Theta_{con}^{(k)}\right) \right. \\
& \left. \times q(V^{(k)}; \Theta_{l}^{(k)}) \right] e^{w_0-w_k}.
\end{aligned}
\end{equation}
\end{small}
It should be noted that deriving the lower bound in Ineq. (\ref{eq: IZY}) and upper bound in Ineq. (\ref{ineq:second_part_final}) enables us to establish an upper limit on the objective function in minimization problem in (\ref{eq:IB}). This makes it easier to minimize by the corresponding loss function during network training, as discussed in \mbox{Sec. \ref{sec: method}-C}.

\subsection{Multi-Frame Correlation Model}
\label{sec: MFCM}
Inspired by the previous work \cite{shao2023task}, we utilize a multi-frame correlation model that leverages variational approximation to capture the temporal dynamics in video sequences. This approach utilizes the temporal redundancy across contiguous frames to model the conditional probability distribution effectively. Our model approximates the next feature in the sequence by considering the variational distribution \( q(Z_{t}^{(k)} | Z_{t-1}^{(k)}, ..., Z_{t-\tau}^{(k)}; \Theta_{\tau}^{(k)}) \), which can be modeled as a Gaussian distribution aimed at mimicking the true conditional distribution of the subsequent frame given the previous frames:
\begin{small}
\begin{equation*}
q\left(Z_{t}^{(k)}|Z_{t-1}^{(k)},...,Z_{t-\tau}^{(k)};\Theta _{\tau}^{(k)}\right)=\mathcal{N} \left( \mu \left( \Theta _{\tau}^{(k)} \right) ,\sigma ^2\left(\Theta _{\tau}^{(k)} \right) \right) ,
\end{equation*}
\end{small}
where \( \mu \) and \( \sigma^2 \) are parametric functions of the preceding frames, encapsulating the temporal dependencies. These functions are modeled using a deep neural network with parameters \( \Theta_{\tau}^{(k)} \) that are learned from data. By optimizing the variational parameters, our model aims to closely match the true distribution, thus encoding the features more efficiently.

\subsection{Network Loss Functions Derivation}\label{Sec: loss}
In this subsection, we design our network loss functions to optimize the information flow in a multi-camera setting according to the IB principle in Sec. \ref{sec: PIB}. 

The first loss function \(\mathcal{L}_1\) aims to minimize the upper bound of the mutual information, following the inequalities derived in (\ref{eq: IZY}) and (\ref{ineq:second_part_final}). \(\mathcal{L}_1\) ensures efficient encoding while preserving essential information for accurate prediction:
\begin{small}
\begin{equation*}
\begin{aligned}
\mathcal{L}_1 =& \sum_{k=1}^K{\underset{\mathrm{The}\ \mathrm{upper} \ \mathrm{bound} \ \mathrm{of} \ -I_w\left( Z^{(k)};Y^{(k)} \right)}{\underbrace{\mathbb{E} [-w_k\log q(Y^{(k)}|Z^{(k)})]}}} + \lambda  \cdot \min \bigg\{ R_{max}, \\
& \underset{\mathrm{The} \ \mathrm{upper} \ \mathrm{bound} \ \mathrm{of} \ I_w\left( X^{(k)};Z^{(k)} \right)}{\underbrace{\mathbb{E} \left[ -\log q(Z^{(k)}|V^{(k)};\Theta _{con}^{(k)})\cdot q(V^{(k)};\Theta _{l}^{(k)}) \right]e^{(w^0-w_k)} }} \bigg\}.
\end{aligned}
\end{equation*}
\end{small}
The first term of \(\mathcal{L}_1\) excludes \(H(Y)\) from Ineq. (\ref{eq: IZY}) because it is a constant. The second term addresses the upper bound of the communication cost required to transmit features from cameras to the edge server. $R_{max}$ is used to clip the over-relaxation for the upper bound, bounding the excessive communications cost, which results in the degradation of training decoder $p(Y^{(k)}|Z^{(k)})$. In Sec. \ref{sec: MFCM}, the Multi-Frame Correlation Model leverages temporal dynamics, which is critical for sequential data processing in video analytics. The second loss function, \(\mathcal{L}^{(k)}_2\), is derived to minimize the KL divergence between the true distribution of frame sequences and the modeled variational distribution:
\begin{equation}
\mathcal{L}_2 =  \sum_{k=1}^{K} D_{KL}\left[p(Z_t^{(k)} | Z_{<t}^{(k)}) || q(Z_t^{(k)} | Z_{<t}^{(k)})\right],
\end{equation}
where $Z_{<t}^{(k)}=(Z_{t-1}^{(k)},...,Z_{t-\tau}^{(k)})$. Given the variability in channel quality and the occurrence of delays, we introduce the third loss function, \(\mathcal{L}^{(k)}_3\), designed to minimize the impact of unreliable data sources while maximizing inference accuracy:
\begin{equation}
\mathcal{L} _3=\sum_{k=1}^K{\left[ 1_{d_{\mathrm{norm},k}<\epsilon}\left( w_k-W_{\mathrm{target}} \right) ^2+1_{d_{\mathrm{norm},k}>\epsilon}\left( w_{k}^{2} \right) \right]},
\end{equation}
where $\epsilon$ denotes a permissible delay that cannot lead to errors in multi-view fusion. $W_{\mathrm{target}}$ represents the target weight for camera without excessive delay. These loss functions collectively aim to optimize the trade-off between data transmission costs and perceptual accuracy, crucial for enhancing the performance of edge analytics in multi-camera systems.

\section{Performance Evaluation}

\subsection{Simulation Setup}
We set up simulations to evaluate our PIB framework, aimed at predicting pedestrian occupancy in urban settings using multiple cameras. These simulations replicate a city environment, with variables like signal frequency and device density affecting the outcomes.

Our simulations use a 2.4 GHz operating frequency, a path loss exponent of 3.5, and a shadowing deviation of 8 dB. Devices emit an interference power of 0.1 Watts, with densities ranging from 10 to 100 devices per 100 square meters, allowing us to test different levels of congestion. The bandwidth is set at 2 MHz, with cameras located about 500 meters from the edge server. We employ the \textit{Wildtrack} dataset from EPFL, which features high-resolution images from seven cameras located in a public area, capturing unscripted pedestrian movements\cite{chavdarova2018wildtrack}. This dataset provides 400 frames per camera at 2 frames per second, documenting over 40,000 bounding boxes that highlight individual movements across more than 300 pedestrians.

The primary measure we use is the multi-object detection accuracy (MODA), which assesses the system’s ability to accurately detect pedestrians based on missed and false detections. We also look at the rate-performance tradeoff to understand how communication overhead affects system performance.

For comparative analysis, we consider three baselines:
\begin{itemize}
    \item \textbf{TOCOM-TEM}\cite{shao2023task}: A task-oriented communication framework using a temporal entropy model for edge video analytics. It leverages the deterministic Information Bottleneck principle to extract and transmit compact, task-relevant features, integrating spatial-temporal data on the server for enhanced inference accuracy.
    \item \textbf{JPEG}\cite{wallace1992jpeg}: A widely adopted image compression standard that reduces the data size of digital images via lossy compression algorithms, commonly used for reducing the communication load in networked camera systems.
    \item \textbf{High Efficiency Video Coding (HEVC)}\cite{bossen2012hevc}: Also known as H.265 and MPEG-H Part 2, this standard provides up to 50\% better data compression than its predecessor AVC (H.264 or MPEG-4 Part 10), maintaining the same video quality, which is critical for efficient data transmission in high-density camera networks.
\end{itemize}

Our code will be available at \href{https://github.com/fangzr/PIB-Prioritized-Information-Bottleneck-Framework}{github.com/fangzr/PIB-Prioritized-Information-Bottleneck-Framework}.

\begin{figure}[t]
  \centering
  \includegraphics[width=0.44\textwidth]{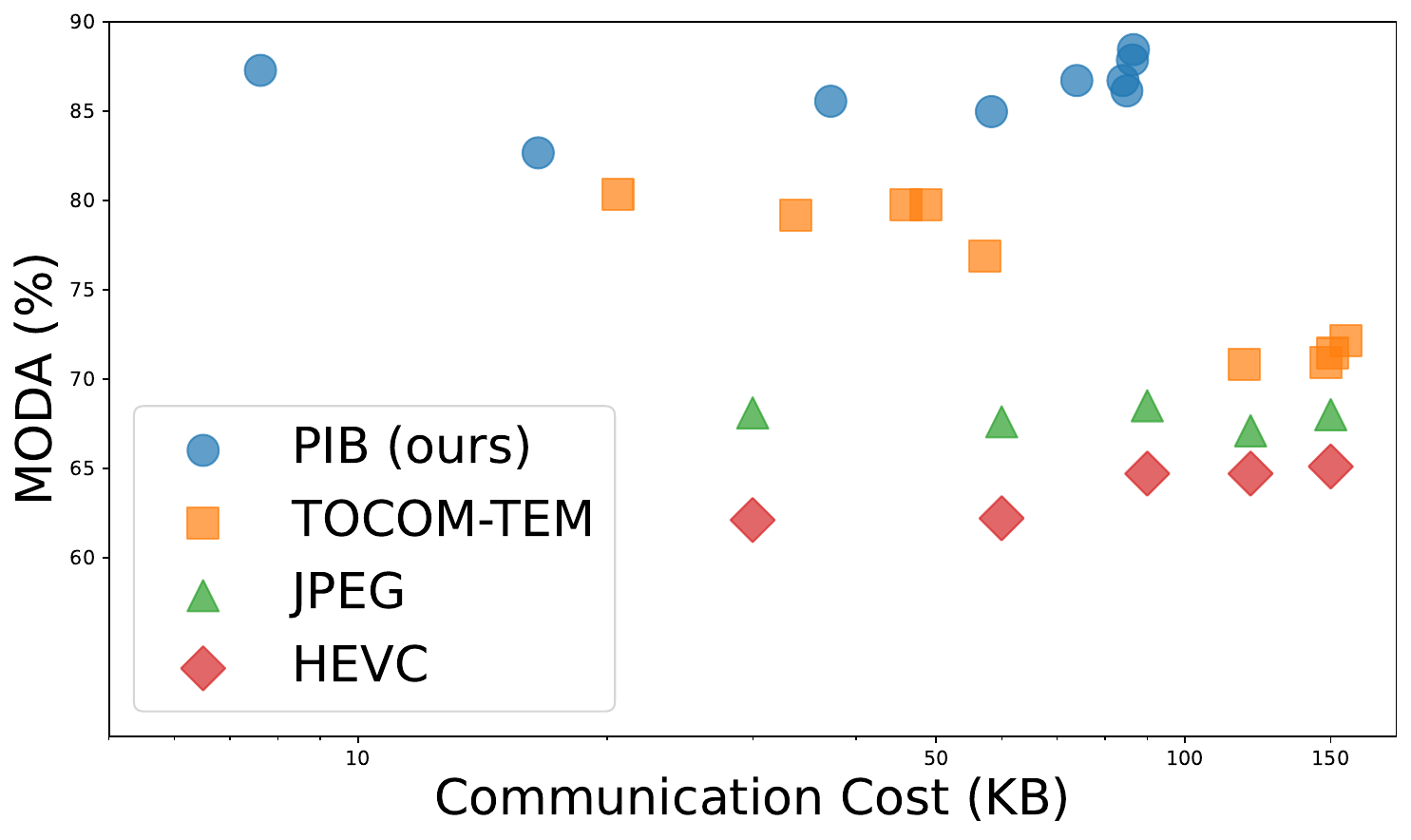}
  \caption{Communication Cost vs MODA.}
  \label{fig:CM}
  \vspace{-3mm}
\end{figure}
In the simulation study, we examine the effectiveness of multiple camera systems in forecasting pedestrian presence. Unlike a single-camera configuration, this method minimizes obstructions commonly found in crowded locations by integrating perspectives from various angles. Nevertheless, this benefit is accompanied by heightened communication overhead. In Fig. \ref{fig:CM}, we observe the relationship between communication costs and MODA, a metric for multi-camera perception. The PIB algorithm exhibits a higher MODA across varying communication costs when compared to TOCOM-TEM, JPEG, and HEVC. This superior performance can be attributed to PIB's strategic fusion of multi-view features, which is informed by both channel quality and the selection of ROI with appropriate priorities. By prioritizing information, PIB effectively mitigates the detrimental effects of delayed information that could potentially degrade the perception accuracy in multi-camera systems.

\begin{figure}[t]
  \centering
  \includegraphics[width=0.44\textwidth]{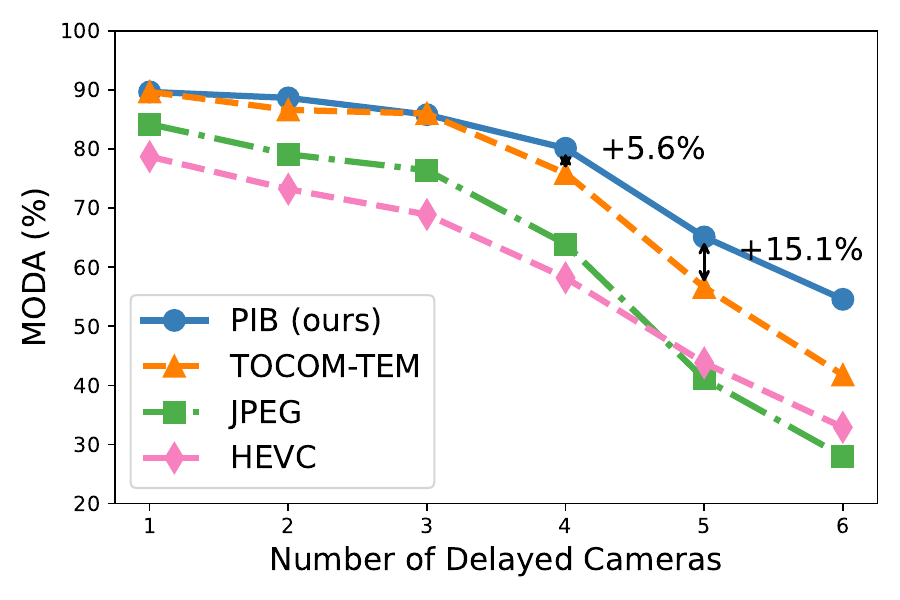}
  \caption{Delayed cameras vs MODA.}
  \label{fig:num-vs-moda}
  \vspace{-3mm}
\end{figure}
Fig. \ref{fig:num-vs-moda} depicts the performance rates of different compression techniques in a multi-view scenario in terms of the number of delayed cameras. Our proposed PIB method and TOCOM-TEM, both utilizing multi-frame correlation models, successfully reduce redundancy across multiple frames, achieving superior MODA at equivalent compression rates. PIB, in particular, utilizes a prioritized IB framework, which technique enables an adaptive balance between compression rate and collaborative sensing accuracy, optimizing MODA across various channel conditions. It is worth noting that JPEG was not consistently outperformed by HEVC compression due to our utilization of the more effective HEIF algorithm derived from HEVC, which inadequately supported the motion prediction module, resulting in compromised performance.

\begin{figure}[t]
  \centering
  \includegraphics[width=0.44\textwidth]{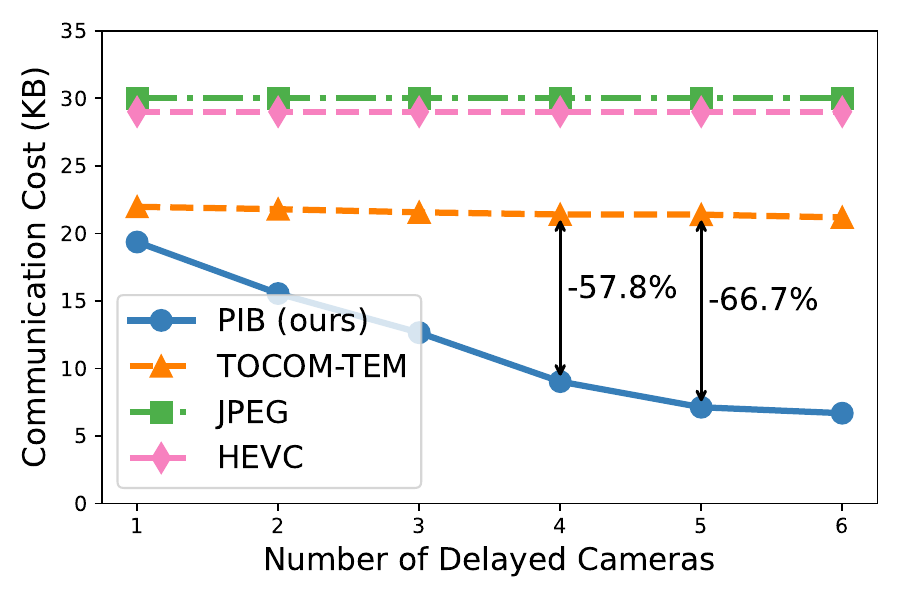}
  \caption{Delayed cameras vs communication cost.}
  \label{fig:num-vs-communication cost}
  \vspace{-3mm}
\end{figure}
In Fig. \ref{fig:num-vs-communication cost}, we analyze the impact of increasing the number of delayed cameras on the communication cost for various algorithms. The PIB algorithm demonstrates a significant reduction in communication costs with a growing amount of delayed cameras. This efficiency is due to the algorithm's priority mechanism that adeptly assigns weights, filtering out the adverse information caused by delays. Consequently, PIB prioritizes the transmission of high-quality features from cameras with more accurate occupancy predictions. When compared to TOCOM-TEM, PIB achieves a remarkable 66.7\% decrease in communication costs while still retaining the precision of multi-camera pedestrian occupancy predictions. For a fair comparison, both JPEG and HEVC methods were set to a uniform compression threshold of 30 KB in this experiment. However, as indicated in Fig. \ref{fig:num-vs-moda}, they have not surpassed the performance of PIB and TOCOM-TEM.

\section{Conclusion}

In this paper, we have proposed the Prioritized Information Bottleneck (PIB) framework as a robust solution for collaborative edge video analytics. Our contributions are two-fold. First, we developed a prioritized inference mechanism to intelligently determine the importance of different camera' FOVs, effectively addressing the constraints imposed by channel capacity and data redundancy. Second, the PIB framework showcases its effectiveness by notably decreasing communication overhead and improving tracking accuracy without requiring video reconstruction at the edge server. Extensive numerical results show that: PIB not only surpasses the performance of conventional methods like TOCOM-TEM, JPEG, and HEVC with a marked improvement of up to 15.1\% in MODA but also achieves a considerable reduction in communication costs by 66.7\%, while retaining low latency and high-quality multi-view sensory data processing under less favorable channel conditions.

\section{Acknowledgement}
This work was supported in part by the Hong Kong SAR Government under the Global STEM Professorship and Research Talent Hub,  the Hong Kong Jockey Club under the Hong Kong JC STEM Lab of Smart City (Ref.: 2023-0108), and the Hong Kong Innovation and Technology Commission under InnoHK Project CIMDA. The work of Y. Deng was supported in part by the National Natural Science Foundation of China under Grant No. 62301300. The work of X. Chen was supported in part by HKU-SCF FinTech Academy R\&D Funding.


%


\bibliographystyle{./IEEEtran}
\bibliography{IEEEabrv,ref}

\end{document}